\documentclass[aps,preprint,floatfix,nofootinbib,showpacs]{revtex4-1}
\pdfoutput=1

\usepackage{graphicx,color,float,amsmath}
\usepackage{hyperref}
\usepackage{MnSymbol}
\usepackage{wasysym}
\usepackage{slashed}
\usepackage{color}

\renewcommand{\thefootnote}{\arabic{footnote}}

\begin{document}

{\small
\begin{flushright}
\end{flushright}}
\vspace*{-2cm}
\title{
Impact of Localization in Early-Universe QCD Phase Transition}

\vspace*{1cm}
\author{
Janus~Capellan~Aban$^{1,2,*}$, Edmayelle~Villavicencio~Alforja$^{3,\dagger}$, Vincent~Gene~L.~Otero$^{3,\ddagger}$}
\affiliation{
\vspace*{.5cm} 
$^1$Department of Mathematics and Physics, University of Santo Tomas, Manila 1008, Philippines
\\ 
$^2$ Research Center for the Natural and Applied Sciences, University of Santo Tomas, Manila 1008, Philippines 
\\
$^3$ Department of Physics, National Taiwan Normal University, Taipei 116, Taiwan
\vspace*{1cm}}


\begingroup
\renewcommand\thefootnote{}\footnote{%
$^*$e-mail: jcaban@ust.edu.ph\vspace{-4pt}\\
$^\dagger$e-mail: edmayelle.alforja@gmail.com\vspace{-4pt}\\
$^\ddagger$e-mail: vglotero@gmail.com}
\addtocounter{footnote}{-1}
\endgroup

\begin{abstract} 
We introduce a phenomenological modification of the MIT bag model equation of state that incorporates quark localization arising from gluon-induced disorder in the quark-gluon plasma. This model effectively reduces the quark degrees of freedom encoded in the product of the disorder activation function $G(W)$ and the localization efficiency factor $H(T)$\,. As a result, the critical temperature is increased roughly by 7\%\,. Employing the Friedmann equation, we find that the onset of the phase transition occurs earlier. Consequently, the mixed phase duration is only $8.22\, \mu s$, which is 24\% shorter than the bag model, and the hadronic phase cools faster. The Stefan-Boltzmann limit is reached at much higher temperatures, causing the energy density and pressure curves of the bag model to shift downward and yielding better agreement with the lattice QCD data from the HotQCD collaboration. Our results show that the localization of quarks plays a significant role in the cooling dynamics of the early universe.

\end{abstract}

\maketitle

\section{Introduction}
In lattice quantum chromodynamics (QCD), it is known that at high temperatures, the quark Dirac eigenmodes can localize near the critical temperature~\cite{Giordano:2013lia,Kovacs:2010wx}. The movement of the localized quarks in this regime is restricted to small spatial regions. In the presence of the disorder induced by the gluon fields background, this phenomenon can be regarded as an Anderson-like localization~\cite{Giordano:2013taa,Giordano:2013kza,Giordano:2014qna,Giordano:2014cua,Giordano:2021qav,Garcia-Garcia:2005azc,Kovacs:2012zq,Horvath:2025ypt,Evers:2007zsx}, in which a transition occurs from delocalized to localized states separated by a mobility edge. Such a localization of quarks may alter the thermodynamic behavior of the quark-gluon plasma (QGP) and change the history of cooling in the early universe.

A well-known description of the equation of state (EOS) of deconfined quarks and gluons is provided by the MIT bag model~\cite{Johnson:1978uy,Chodos:1974je,Chodos:1974pn,DeGrand:1975cf}. In the confinement phase, this model treats the quarks as non-interacting particles contained within a bag. This bag acts like an infinite potential well in which quarks are free to move but are prohibited from escaping. It has an associated bag constant $B$ that encapsulates the vacuum pressure, maintains the quark confinement, and counteracts the outward pressure they exert on the bag wall. When the outward pressure exceeds the inward vacuum pressure, the bag becomes unstable and, if this condition persists, the QGP phase forms eventually. As a result, the quarks can move over large distances and contribute to thermodynamical quantities such as energy density $\varepsilon$ and pressure $p$\,. Although the bag model reproduces many physical features of the QGP, it treats quarks as non-interacting particles and neglects disorder-induced localization. The recent lattice results~\cite{Giordano:2013lia,Kovacs:2010wx,Giordano:2013taa,Giordano:2013kza,Giordano:2014qna,Giordano:2014cua,Giordano:2021qav,Garcia-Garcia:2005azc,Kovacs:2012zq,Horvath:2025ypt,Evers:2007zsx} on localization motivate us to extend this bag framework by including the disorder and temperature dependent suppression of quark degrees of freedom (d.o.f.)\,. 

In this study, we examine the localization effects of the disorder $W$ arising from gluon fields on the EOS of the MIT bag model.  In particular, we apply this framework to the early-universe QCD phase transition, where the localization of quarks is analogous to the Anderson localization. In effect, the d.o.f. of quarks will be suppressed due to the limitation of their motion. The relevant suppression factor $F_{ext}(W,T)$ is expressed as a function of the product of the disorder activation function $G(W)$ and the localization efficiency factor $H(T)$. For simplicity, we assume that the disorder parameter $W$ is directly proportional to the temperature $T$. Moreover, the suppression factor reduces the quark contribution to the energy density and pressure in the early-universe scenario. 

We assume that the universe is homogeneous and isotropic~\cite{Weinberg:1972} as it expands from the onset of the big bang. This makes the time evolution of thermodynamical quantities, such as energy density and pressure to be governed by the Friedmann equation  
~\cite{Guardo:2014rta,Florkowski:2010mc,Ornik:1987up} 
\begin{align}
\label{eq:Friedmannequationfirst}
-\frac{d\varepsilon}{3\sqrt{\varepsilon (T)}(\varepsilon(T) + p(T))}
= \sqrt{\frac{8\pi G}{3}} dt \:\:. 
\end{align}
Using the above Eq.~\eqref{eq:Friedmannequationfirst}, one can obtain the time evolution of temperature $T(t)$ to examine the impact of the quark localization on the cooling of both plasma and hadron phases. Consequently, the localization alters the duration of the mixed-phase regime. To assess our model, we compare our results with lattice QCD data from the HotQCD collaboration~\cite{HotQCD:2014kol}. Indeed, the localization model yields better agreement than the MIT bag model, using conditions consistent with a realistic QCD EOS applied to relativistic heavy-ion collisions at RHIC~\cite{Florkowski:2010mc}. 

This work is organized as follows: In Section~\ref{section:MITbagmodel}, we briefly discuss the MIT bag model EOS~\cite{Chodos:1974je,Fogaca:2009wf} including the contribution from the electroweak sector. In Section~\ref{Section:ModifiedMITBagModelwithLocalization}, we introduce the localization effects of the disorder arising from the gluon fields as a phenomenological modification of the MIT bag model. In this localization framework, we consider a suppression factor $F_{ext}(W,T)$ for the d.o.f. of quarks defined as a function of the product of the disorder activation function $G(W)$ and the localization efficiency factor $H(T)$. We then apply this localization model to the early-universe QCD phase transition, where the cosmic expansion is governed by the Friedmann equation~\cite{Guardo:2014rta,Florkowski:2010mc,Ornik:1987up}. For Section~\ref{section:resultsanddiscussion}, we discuss the implications of quark localization on the thermodynamical quantities such as energy density and pressure, and its consequences to the cooling dynamics of the plasma and hadronic phases in the early-universe scenario. We also compare our findings with lattice QCD data from the HotQCD collaboration~\cite{HotQCD:2014kol}. Finally, our conclusion is presented in Section~\ref{section:conclusion}. 

\section{The MIT bag Model}
\label{section:MITbagmodel}
In this section, we consider the EOS of the QGP as described by the well-known MIT bag model~\cite{Chodos:1974je}. In this model, quarks and gluons are free to move within a finite region called the bag. However, once the outward pressure on the bag wall exceeds the confining pressure, quarks and gluons can move across larger spatial regions, indicating the deconfined phase of the system. As described by the MIT bag model for temperature $T>T_c$, the energy density and pressure in the deconfined phase are given by~\cite{Chodos:1974je,Fogaca:2009wf} 
   \begin{align}
    \varepsilon_{qgp}(T)=g_{qgp}\dfrac{\pi^2}{30}T^4 + B \:\:\:\:\:\textrm{and}\:\:\:\:\: p_{qgp}(T)=g_{qgp}\dfrac{\pi^2}{90}T^4 - B,
\end{align}
while for  $T<T_c$ we have
\begin{align}
    \varepsilon_\pi (T)=g_{\pi}\dfrac{\pi^2}{30}T^4\:\:\:\:\:\textrm{and}\:\:\:\:\:
    p_\pi(T)=g_{\pi}\dfrac{\pi^2}{90}T^4,
\end{align}
where $T_c$ is the critical temperature for the phase transition. The quantities $g_{qgp}=g_g + \frac{7}{8}g_q$ and $g_{\pi}$ denote the d.o.f. of QGP and the pion gas, respectively. Here, the value of $g_{\pi}$ is 3, while for two (three) different quark flavors, $g_{qgp}=37$ (47.5). Interestingly, for very high temperatures, the bag EOS approaches the Stefan-Boltzmann limit 
\begin{align}
\dfrac{\varepsilon_{qgp}}{T^4}=\dfrac{37 }{30} \pi^2\:\:\:\textrm{and}\:\:\: \dfrac{\varepsilon_{qgp}}{T^4}=\dfrac{47.5 }{30}\pi^2    
\end{align}
for two and three quark flavors, respectively.

The inclusion of the electroweak sector will generate additional contributions to the energy density and pressure given by  
\begin{align}
    \varepsilon_{ew} (T)=g_{ew}\dfrac{\pi^2}{30}T^4
\:\:\:\:\:\textrm{and}\:\:\:\:\:
    p_{ew}(T)=g_{ew}\dfrac{\pi^2}{90}T^4
\end{align}
which are relevant within the temperature range considered in this study as in~\cite{Florkowski:2010mc}. By treating electroweak particles as massless, the effective d.o.f. becomes $g_{ew}=14.25$~\cite{Yagi:2005yb}, coming from the photon along with different leptonic particles, including electrons, muons, three neutrino flavors, and their associated antiparticles, respectively. Hence, the resulting energy density and pressure with contribution from the electroweak sector modify the bag model EOS for $T>T_c$, yielding
\begin{align}
    \varepsilon_B (T)=g_1\dfrac{\pi^2}{30}T^4 + B 
    \:\:\:\:\:\textrm{and}\:\:\:\:\:
    p_B(T)=g_1\dfrac{\pi^2}{90}T^4 - B,
\end{align}
and for $T<T_c$
\begin{align}
\label{eq:piongaseos}
    \varepsilon_{\pi,ew} (T)=g_2\dfrac{\pi^2}{30}T^4 \:\:\:\:\:\textrm{and}\:\:\:\:\:
    p_{\pi,ew}(T)=g_2\dfrac{\pi^2}{90}T^4,
\end{align}
where $g_1=g_{qgp}+g_{ew}$ and $g_2=g_{\pi}+g_{ew}$\,. 

The critical temperature $T_c$ can be determined by equating the pressures in the two regimes, that is
\begin{equation}
    g_1\dfrac{\pi^2}{90}T_c^4 - B=g_2\dfrac{\pi^2}{90}T_c^4 
\end{equation}
which consequently gives
\begin{align}
\label{eq:qgp+ew:criticaltemp}
    T_c= \left[\dfrac{90B}{\pi^2 (g_1 - g_2 )}\right]^{1/4}\,. 
\end{align}
Note that the denominator of Eq.~\eqref{eq:qgp+ew:criticaltemp} is equal to the difference $g_{qgp}-g_\pi$. This implies that the critical temperature obtained from the MIT bag model will remain unchanged even if there is a contribution from the electroweak sector. In particular, for two quark flavors with bag constants $B=(165\, \textrm{MeV})^4$ and $B=(235\, \textrm{MeV})^4$, the corresponding critical temperatures are $T_c=119\,\textrm{MeV}$ and $T_c=169\,\textrm{MeV}$, respectively. Notably, the two thermodynamic quantities satisfy the relation
\begin{align}
\label{eq:pressureanddensityrelation}    
p_B [\varepsilon_B(t)]=\dfrac{1}{3}[\varepsilon_B(t)-4B],
\end{align}
where both quantities are considered to evolve with time $t$. A special case of the bag model happens when only gluons participate thermodynamically~\cite{Chodos:1974je,Fogaca:2009wf}, corresponding to the reduced energy density and pressure
\begin{align}
    \varepsilon_g (T)=\dfrac{8 \pi^2}{15}T^4 + B 
    \:\:\:\:\:\textrm{and}\:\:\:\:\:
    p_g(T)=\dfrac{8 \pi^2}{45}T^4 - B\:.
\end{align}

\section{Modified MIT Bag Model with Localization}
\label{Section:ModifiedMITBagModelwithLocalization}
It is important to note that the QGP near the critical temperature $T_c$ remains a strongly interacting system~\cite{PHENIX:2003qra,STAR:2000ekf} as evident by the observed elliptic flow measurements from the relativistic heavy ion collider (RHIC). As inferred from the Shapiro delay of the binary milliseconds pulsar PSR J1614-2230~\cite{Kalam:2016mmr}, the possible existence of quark stars made up of cold QGP with masses on the order of one solar mass suggests that the MIT bag model should be modified to account for the interaction among quarks. Several works proposing modifications to the MIT bag model can be found in~\cite{Sanches:2014gfa,Alford:2004pf,Giacosa:2010vz,Begun:2010eh,Begun:2012zz}. For instance, this study~\cite{Begun:2012zz} introduced additional linear and quadratic temperature-dependent terms in both the energy density and pressure, resulting in a reduction of the Stefan-Boltzmann limit. In relation to localization, the eigenstates of the Dirac operator, known as quark eigenmodes, have the ability to be localized near the critical temperature~\cite{Giordano:2013lia}, where the phase transition occurs. This mechanism can be referred to as Anderson-like localization, where the quark wavefunction is spatially constrained according to its localization length~\cite{Evers:2007zsx}\,. A more detailed discussion about quark localization can be found in~\cite{Giordano:2013lia,Giordano:2013taa,Giordano:2013kza,Giordano:2014qna,Giordano:2014cua,Giordano:2021qav,Garcia-Garcia:2005azc,Kovacs:2012zq,Horvath:2025ypt}. 

We now consider a phenomenological modification of the MIT bag model EOS in which the gluons are fully deconfined, while the quarks can be localized due to disorder in the gluon plasma. This leads to a reduced number of quarks participating thermodynamically, weighted by the extended-state fraction $F_{ext}(W,T)$, yielding an effective number of quark d.o.f. that depends on disorder $W$ and temperature $T$
\begin{align}
    g_{q}^{eff}=F_{ext}(T,W) g_q,
\end{align}
where $g_q=24\, (36)$ for two (three) quark flavors. The extended-state fraction is defined as
\begin{align}
  F_{ext}(T,W)=\left[1-F_{loc}(W,T)\right] 
\end{align}
such that the localized-state fraction  $F_{loc}(W,T)$ is given by
\begin{align}
    F_{loc}(W,T)=F_{max}\cdot G(W)\cdot H(T),
\end{align}
where $G(W)$ denotes the disorder activation function and $H(T)$ represents the localization efficiency factor with a cap factor $F_{max}$. The parameter $F_{max}$ indicates the maximum allowable fraction of localized quarks. We assume analytical profiles for the two functions given by
\begin{align}
\label{eq:analfunctions}
  G(W)=  1-\exp\left[-\left(\frac{W}{W_\times}\right)^q\right]\:\:\:\textrm{and}\:\:\:H(T)=\dfrac{1}{1+\left(\dfrac{T}{T_*}\right)^p}
\end{align}
with positive constants $p$ and $q$. For the disorder activation function $G(W)$, we further assume that the disorder parameter $W$ is directly proportional to the temperature which means that an increase in temperature results to a more disordered gluon plasma. This assumption gives the relation
\begin{align}
\dfrac{W}{W_\times}=\dfrac{T}{T_\times},
\end{align}
implying that the disorder activation function can be rewritten explicitly as a function of temperature
\begin{align}
\label{eq:disorderactivation}
  G(T)=  1-\exp\left[-\left(\frac{T}{T_\times}\right)^q\right]\:\:\:.
\end{align}
Note that as the temperature $T\rightarrow 0$,  the disorder activation function approaches $G(T )\rightarrow 0$, indicating that the disorder is very weakly activated in this regime. Conversely, as temperature $T\rightarrow+\infty$,  the localization efficiency factor vanishes $H(T)\rightarrow 0$, implying that the quarks can freely participate thermodynamically with negligible localization effects. Hence, the analytical functions $G(T)$ and $H(T)$ considered here are bounded, well-behaved, and satisfy the expected physical limits across the entire temperature range. 

The temperatures $T_\times$ and $T_*$ depicted in Eqs.~\eqref{eq:analfunctions} and \eqref{eq:disorderactivation} refer to the characteristic temperature scales associated with disorder and localization, respectively. Under these assumptions, the energy density and pressure of the modified MIT bag model EOS with localization for temperatures $T>T_c^{loc}$ can be written as 
\begin{align}
\label{eq:localizationenergydensityandpressure}
 \varepsilon_{loc} (T)= \dfrac{\pi^2}{30}\left[g_{g} + g_{ew}+ \dfrac{7}{8} g_{q}^{eff} \right]T^4 + B\\
  p_{loc} (T)= \dfrac{\pi^2}{90}\left[g_{g} + g_{ew} + \dfrac{7}{8} g_{q}^{eff}   \right]T^4 - B,
\end{align}
where we include the electroweak sector such that the phase transition occurs at the critical temperature $T_c^{loc}$. For the regime $T<T_c^{loc}$, the energy density and pressure are still described by the massless pion gas given in Eq.~\eqref{eq:piongaseos}. Again, the critical temperature $T_c^{loc}$ occurs when the pressure in both regimes is equal, that is 
\begin{align}
\label{eq:criticaltemploc}
    p_{loc}(T_c^{loc})=p_{\pi,ew} (T_c^{loc})\:\:. 
\end{align}
From the above Eq.~\eqref{eq:criticaltemploc}, the expression for the critical temperature is obtained as
\begin{align}
\label{eq:localozationtemperature}
    T_c^{loc} =\left[ \dfrac{90B}{\pi^2 \left(13+ \dfrac{7}{8} \left(g_{q}^{eff} \right)  \right) }\right]^{1/4}\:.
\end{align}
Let us consider the time evolution of our thermodynamic quantities. We assume that our universe is homogeneous and isotropic~\cite{Weinberg:1972}, then the Einstein's field equations reduce to the Friedmann equation~\cite{Guardo:2014rta,Florkowski:2010mc,Ornik:1987up} 
\begin{align}
\label{eq:Friedmannequation}
-\frac{d\varepsilon}{3\sqrt{\varepsilon}(\varepsilon + p)}
= \sqrt{\frac{8\pi G}{3}} dt
\end{align}
as derived using the Friedmann-Lemaître-Robertson-Walker (FLRW) metric. Note that in the localization model, the pressure can be expressed as a function of the energy density for $T>T_c^{loc}$
\begin{align} 
p_{loc} [\varepsilon_{loc}(t)]=\dfrac{1}{3}[\varepsilon_{loc}(t)-4B]\,
\end{align}
which resembles Eq.~\eqref{eq:pressureanddensityrelation}. Using Eq.~\eqref{eq:Friedmannequation} we obtain the differential equation
\begin{align}
\label{eq:Friedmanneq}
    -\frac{d\varepsilon_{loc}}{4\sqrt{\varepsilon_{loc}}\,\left( 
    \varepsilon_{loc} -B
    \right)} = \chi \, dt,
\end{align}
where $\chi=\sqrt{\frac{8\pi G}{3}}$. By solving it analytically, we can have
\begin{align}
    \dfrac{\tanh^{-1}
    (\sqrt{\dfrac{\varepsilon_{loc}}{B}})}{2\sqrt{B}}\Bigg|_{+\infty}^{\varepsilon_{loc}(t) }= \chi \,t,
\end{align}
where the limits of integration on the right-hand side of Eq.~\eqref{eq:Friedmanneq} correspond to time $t=0$ up to time $t$, which are associated to energy density $\varepsilon_{loc}=+\infty$ and $\varepsilon_{loc}(t)$, respectively.

By rewriting the inverse hyperbolic tangent in its natural logarithm form
\begin{align}
    \ln\left( \dfrac{r-1}{r+1}   \right) \Bigg|_{\infty}^{\varepsilon_{loc}(t) } = -4\sqrt{B}\,\chi\, t
\end{align}
which implies that 
\begin{align}
\label{eq:naturallogarithmform}
\left( \dfrac{r-1}{r+1}   \right) = \exp\left[-2(2\sqrt{B}\, \chi \,t) \right]
\end{align}
with  $r=\sqrt{\dfrac{\varepsilon_{loc}(t)}{B}}$. As time $t$ approaches zero, the energy density becomes extremely large due to the very high temperature, making the corresponding quantity $r$ large as well. This yields the equation 
\begin{align}
\label{eq:naturallogarithmformconstant1}
\left( \dfrac{r-1}{r+1}   \right) = \exp\left(-2x \right)\,
\end{align}
such that $x=2\sqrt{B}\, \chi \,t$\,. It is important to note that the function $r=\coth x$ satisfies Eq.~\eqref{eq:naturallogarithmformconstant1}\,. Therefore, the time evolution of energy density will be
\begin{align}
\label{eq:energydensityevolutionqgploc}
   & \sqrt{\dfrac{\varepsilon_{loc}(t)}{B}} = \coth (x) \\
   \implies & \varepsilon_{loc}(t)=B\coth^2\left( \dfrac{t}{t_U}\right)
\end{align}
 with time scale parameter $t_U=\sqrt{\dfrac{3}{32\pi GB}}$. In particular, for the bag constants 
\begin{align}
B=(165 \,\textrm{MeV})^4\:\:\:\textrm{and}\:\:\:B=(235 \,\textrm{MeV})^4,
\end{align}
the corresponding values of the time-scale parameter are $t_U=51\, \mu s$ and  $t_U=25\, \mu s$~\cite{Florkowski:2010mc}, respectively. The time evolution of the energy density in Eq.~\eqref{eq:energydensityevolutionqgploc} is valid only within the interval $0<t\leq t_1$, where $t_1$ satisfies the equation 
\begin{equation}
\label{eq:t1localization}
    \varepsilon_{loc}(t_1)=B\coth^2\left( \dfrac{t_1}{t_U}\right)=g^{eff} (T_c^{loc})\dfrac{\pi^2}{30}(T_c^{loc})^4 + B\:
\end{equation}
with $g^{eff}=g_{g} + g_{ew}+ \dfrac{7}{8} g_{q}^{eff}$
The time $t_1$ marks the onset of the phase transition from QGP to the hadronic phase. In the case of the MIT bag model with the inclusion of the electroweak sector, the bag constants 
\begin{align}
B=(165 \,\textrm{MeV})^4\:\:\:\textrm{and}\:\:\:B=(235 \,\textrm{MeV})^4 
\end{align}  
correspond to the $t_1=23\,\mu s$ and  $t_1=11.4\,\mu s$, respectively, for two quark flavors. The final moment of the phase transition is denoted by the time $t_2$ satisfying the relation~\cite{Yagi:2005yb,Florkowski:2010mc} 
\begin{align}
\label{eq:timeintervalphasetransition}
t_{2} - t_{1} = \frac{4 t_{U}}{3} \sqrt{s - 1} 
\left[ \arctan\!\left(\sqrt{4s - 1}\,\right) - \arctan\!\left(\sqrt{3}\,\right) \right]
\end{align}
such that $s$ denotes the ratio of the d.o.f. in the plasma and hadronic phases, given in our localization model by $s=\dfrac{g^{eff}}{g_2}$. In the case of the bag model, the calculated values of the time $t_2$ for the bag constants 
\begin{align}
B=(165 \,\textrm{MeV})^4\:\:\:\textrm{and}\:\:\:B=(235 \,\textrm{MeV})^4 
\end{align}  
are $t_2=44.93\,\mu s$ and $t_2=22.2 \,\mu s$~\cite{Florkowski:2010mc}, respectively.  The time evolution of the hadronic phase obtained from Eq.~\eqref{eq:piongaseos} can be derived similarly, starting from the Friedmann equation to arrive at 
\begin{align}
\label{eq:timeevolutionhadronloc}
    \dfrac{1}{\sqrt{\varepsilon_{\pi,ew}}}\Bigg\vert_{\varepsilon_{\pi,ew}(t_2)}^{\varepsilon_{\pi,ew}(t)}=2\sqrt{\dfrac{8\pi G}{3}}t\Bigg \vert_{t=t_2}^{t}\,.
\end{align}
This simplifies into
\begin{align}
 \dfrac{1}{\sqrt{a_{\pi,ew} T^4}}-2\,\chi\, t = \dfrac{1}{\sqrt{a_{\pi,ew} (T_{loc}^{c})^4}}-2\,\chi\, t_2 
\end{align}
with  $a_{\pi,ew}=(g_{\pi}+g_{ew})\dfrac{\pi^2}{30}$\,.

\section{Results and Discussion}
\label{section:resultsanddiscussion}
\subsection{Localization due to disorder}
Note that the effective quark d.o.f. under suppression is given by $g_q^{\mathrm{eff}}=F_{ext}(T,W)g_q$ 
such that $F_{ext}(T,W)=(1-F_{loc}(W,T))$. The effect of disorder $W$ caused by the gluon fields in localizing quarks is encoded in $F_{loc}(W,T)$, which is also a function of temperature $T$. In our case, we assume that the disorder dependence $W$ in
\begin{align}
G(W)=1-\exp\left[-(\frac{W}{W_\times})^q\right]
\end{align}
is directly proportional to temperature $T$, where $W_\times$ is a reference disorder parameter. That is, as the temperature increases, the gluon fields become more chaotic, leading to more disorder. With this assumption, the disorder activation function $G(W)$ can be expressed as a temperature-dependent function
\begin{align}
\label{disorderactivation}
  G(T)=1-\exp\left[-(\frac{T}{T_\times})^q\right],
\end{align}
where $q$ is a positive constant. This means that an increase in disorder or temperature will make the factor $G(T)$ larger, thereby making $F_{ext}(T)$ smaller. The characteristic temperature scale $T_\times$ indicates the transition point at which the disorder activation function $G(T)$ changes from being small to significant. The exponent $q$ controls how rapidly the disorder activation function rises from 0 to 1 as the temperature increases. Observe that when $T<<T_\times$, or equivalently $\dfrac{T}{T_\times}<<1$, we have
\begin{align}
  \exp\left[-(\frac{T}{T_\times })^q\right] \approx 1-(\frac{T}{T_\times})^q,  
\end{align}
yielding the approximate disorder activation function
\begin{align}
\label{eq:disorederactivationapprox}
  G(T)\approx 1-\left[1-(\frac{T}{T_\times})^q\right]=(\frac{T}{T_\times})^q
\end{align}
which is very tiny in this regime. However for $T\approx T_\times$, we obtain $G(T)\approx (1-e^{-1})$ which is about $63 \%$ of the maximum disorder activation. In the limiting case $T>> T_\times$, $G(T)\approx 1$, indicating that the disorder activation is fully switched on. Next, we examine the  localization efficiency factor $H(T)$, which depends explicitly on temperature 
\begin{align}
\label{eq:localizationfactor}
    H(T)= \dfrac{1}{1+\left(\dfrac{T}{T_*}\right)^p},
\end{align}
where $T_*$ is the associated characteristic temperature scale. The temperature $T_*$ marks the transition point beyond which localization effects become less pronounced. The factor $H(T)$ decreases with increasing temperature. Observe that when $T=T_*$, $H(T)\approx \frac{1}{2}$, corresponding to half of its maximum value. This indicates a transition from strong to weak localization. In the limit $T<< T_*$, $H(T)\approx 1$, which means that the localization caused by disorder is highly efficient and at its maximum, signifying very strong localization. For very high temperatures $T>>T_*$, or equivalently $\dfrac{T}{T_*}>>1$, Eq.~\eqref{eq:localizationfactor} gives
\begin{align}
    H(T)\approx \left(\dfrac{T_*}{T}\right)^p\,.
\end{align}
Hence for  $T>>T_*$, $H(T)\approx \left(\dfrac{T_*}{T}\right)^p$, which corresponds to very weak localization. In this regime, the localization is controlled by $\dfrac{1}{T^p}$, where $p$ dictates the steepness of the decline in localization efficiency $H(T)$ as temperature $T$ increases.

\begin{figure}[htbp]
    \centering
    \includegraphics[width=8.1cm]{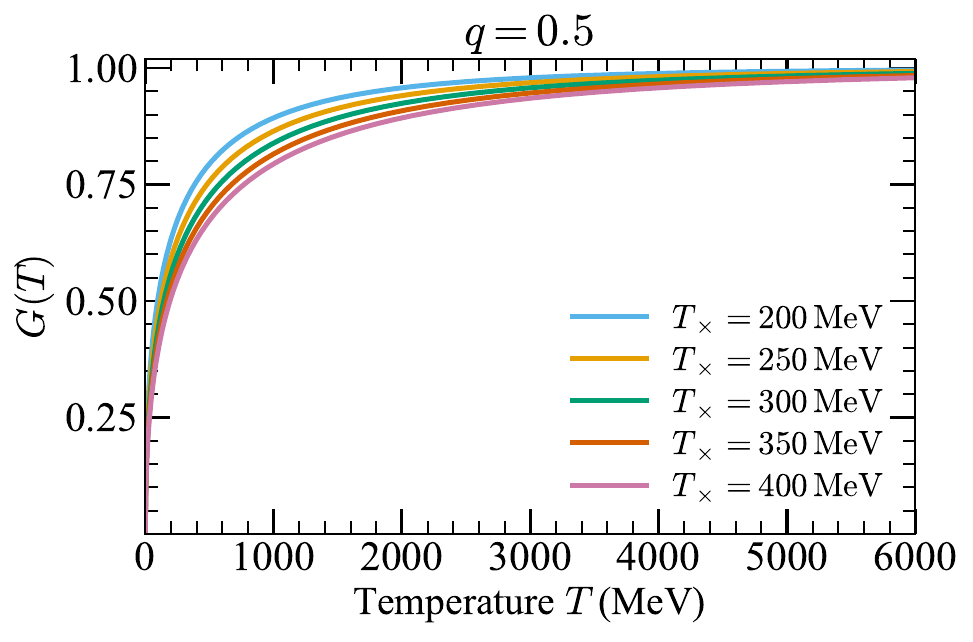}
    \includegraphics[width=8.1cm]{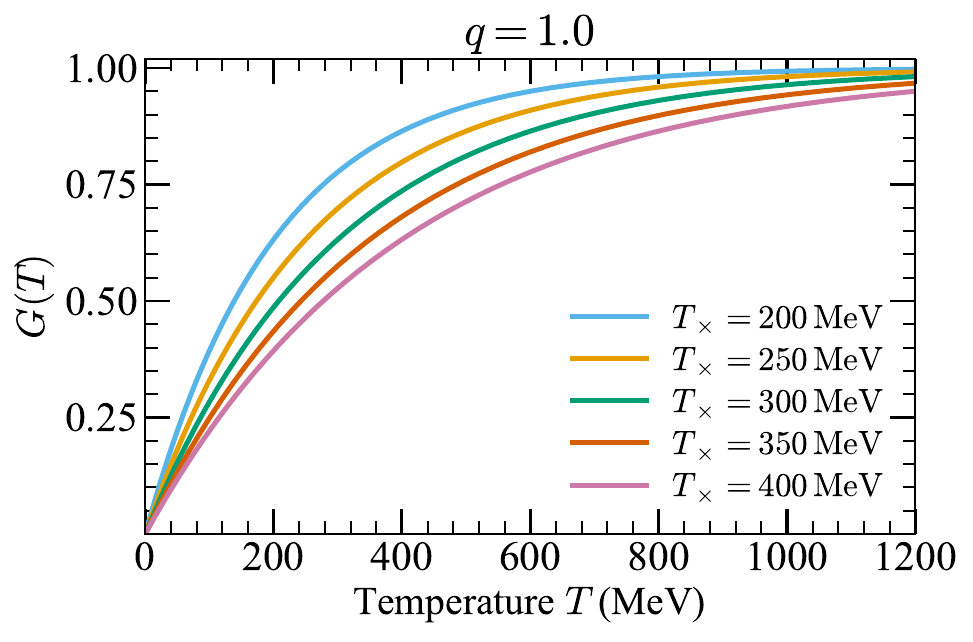} \\
     \includegraphics[width=8.1cm] {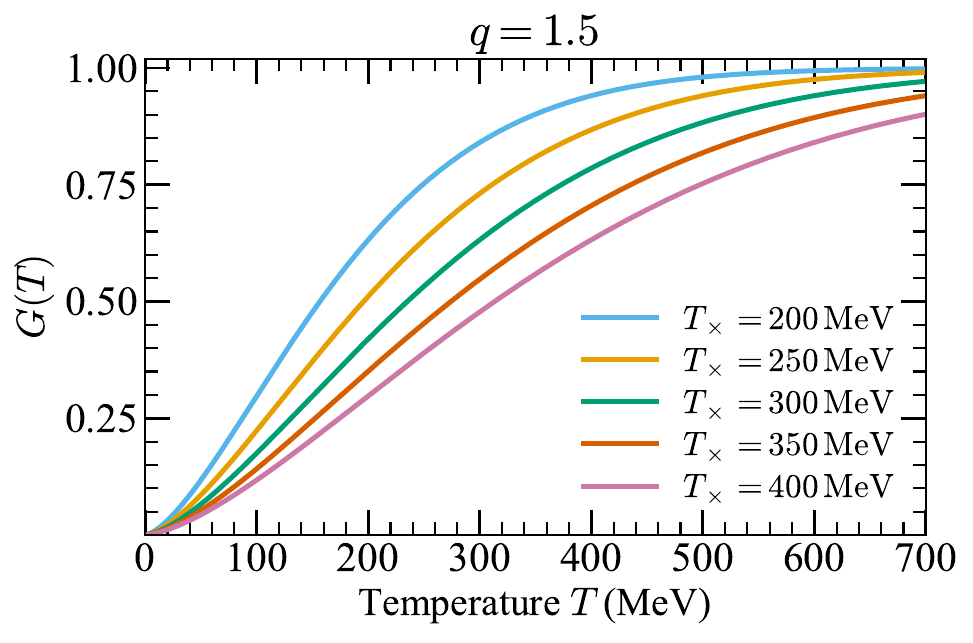}
    \includegraphics[width=8.1cm]{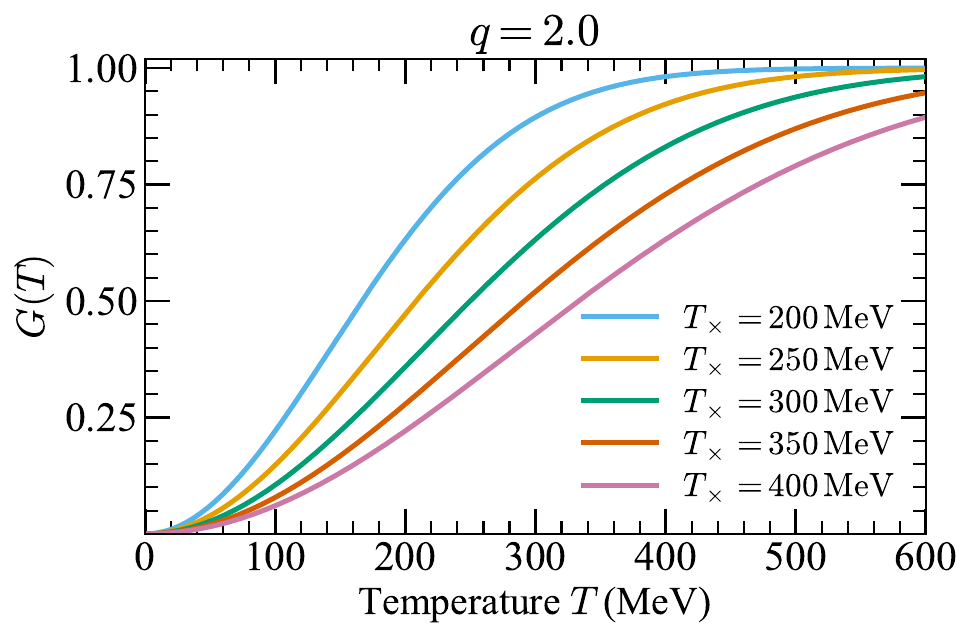}
    \caption{ Plots of the disorder activation function $G(T)$ for varying characteristic temperature scales $T_\times=$ 200 MeV, 250 MeV, 300 MeV, 350 MeV, and 400 MeV at different exponent values $q=0.5, 1.0, 1.5$, and 2.0\,.  }
    \label{fig:disorderact} 
\end{figure}

From our analysis, the maximum possible value of the product $G(T)H(T)$ is 1, which occurs only when $T>>T_\times$ and $T<< T_*$, respectively. This scenario happens in the temperature range
\begin{align}
\label{eq:maximalcondition}
  T_\times << T << T_*
\end{align}
in which the presence of the disorder can localize the quarks most efficiently. In Fig.~\ref{fig:disorderact}
the plots show the disorder activation function $G(T)$ for different characteristic temperature scales $T_\times=$ 200 MeV, 250 MeV, 300 MeV, 350 MeV and 400 MeV. It is important to note that the disorder activation function $G(T)$ increases with temperature. Notably, for a fixed temperature $T$ and exponent value $q$, $G(T)$ increases as the characteristic temperature scale $T_\times$ decreases. This behavior is reflected in the exponential term $\exp\left[-( \dfrac{T}{T_\times})^q\right]$ of Eq.~\eqref{disorderactivation} which becomes smaller for lower values $T_\times$ at a fixed temperature $T$. As a result, the disorder activation function $G(T)$ becomes larger for lower values of the characteristic temperature scale $T_\times$. It also saturates  earlier for higher values of $q$ since the ratio $\left(\dfrac{T}{T_\times}\right)^q$ increases more rapidly for $T>T_\times$ as $q$ increases, causing $\exp\left[-\left( \dfrac{T}{T_\times}\right)^q\right]$ to approach zero more quickly. This means that the maximum activation of disorder is reached at relatively lower temperatures. However for $T<T_\times$, the function $G(T)$ is larger for smaller values of $q$, since the ratio $\dfrac{T}{T_\times}<1$  implies a faster decay of $\left(\dfrac{T}{T_\times }\right)^q$ for higher values of $q$. Hence, for $T<T_\times $, the growth of disorder activation is favored by lower values of $q$. At temperature $T=T_\times $, the amount of disorder activation function $G(T)$ is 0.63\,.

\begin{figure}[htbp]
    \centering
    \includegraphics[width=8.1cm]{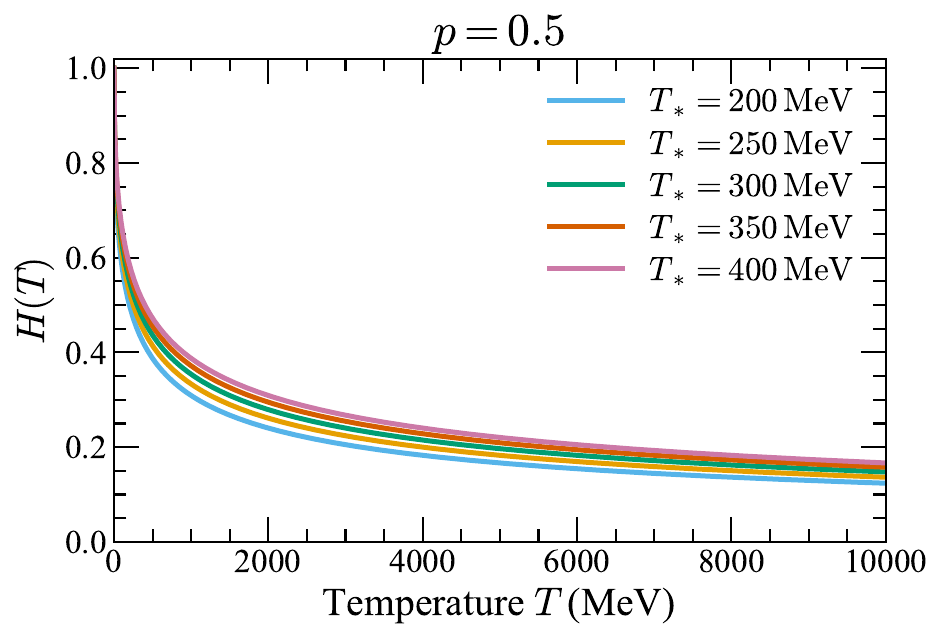}
    \includegraphics[width=8.1cm]{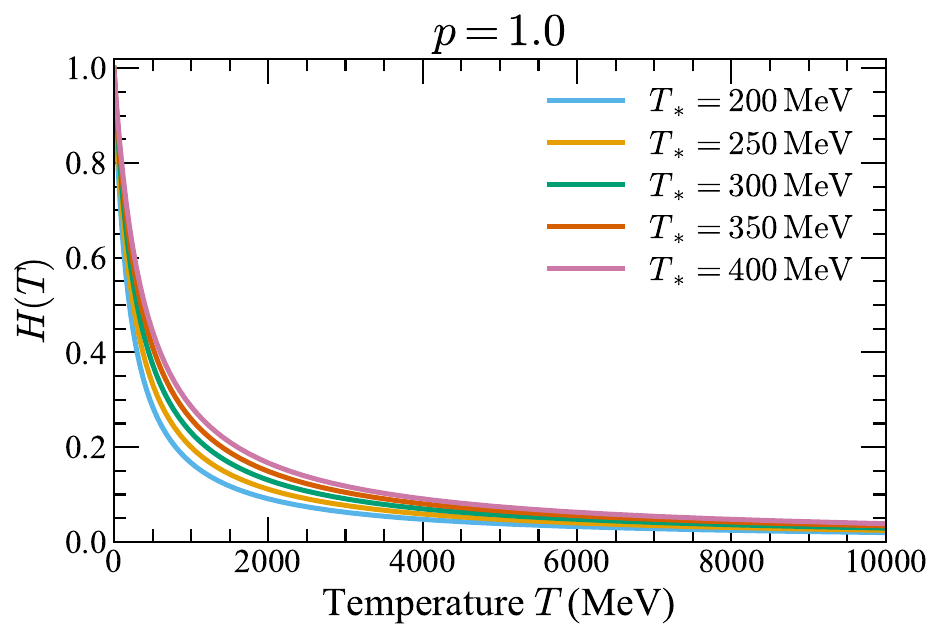} \\
     \includegraphics[width=8.1cm] {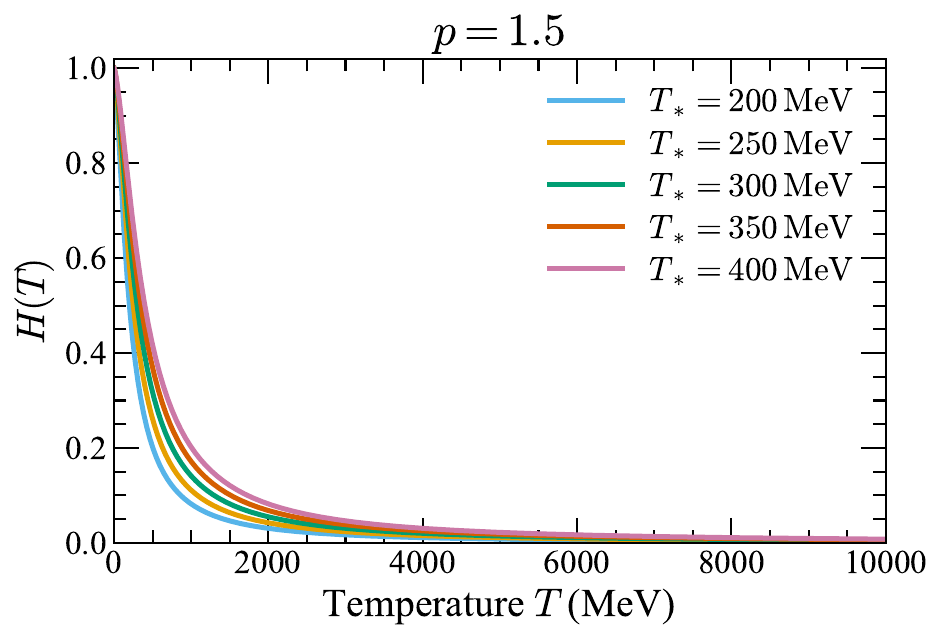}
    \includegraphics[width=8.1cm]{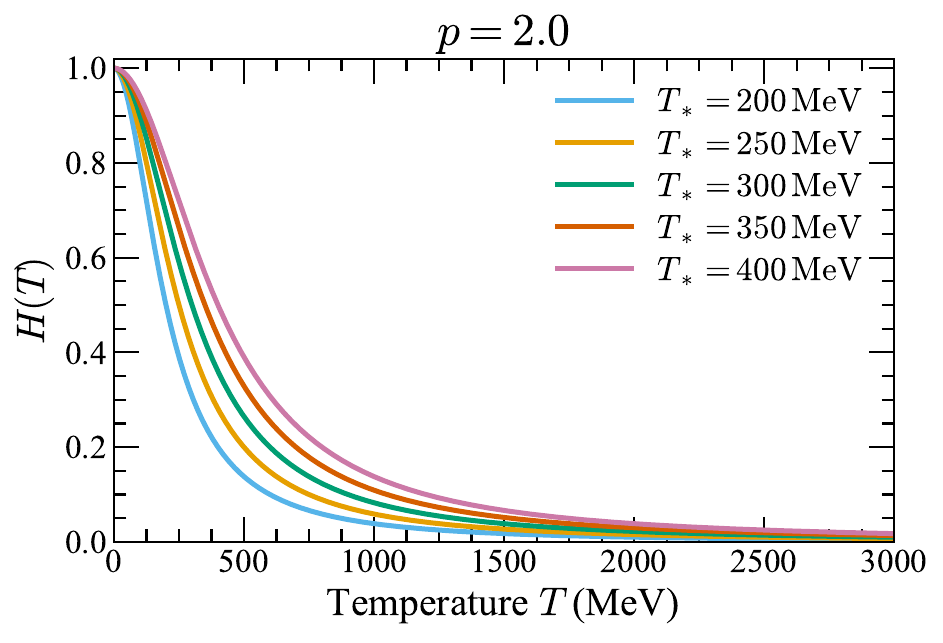}
    \caption{ The plots of the localization efficiency factor $H(T)$ as a function of temperature for varying characteristic temperature scales $T_*=$ 200 MeV, 250 MeV, 300 MeV, 350 MeV, and 400 MeV at different exponent values $p=0.5, 1.0, 1.5$, and $2.0$\,. }
    \label{fig:loceff} 
\end{figure}

Now, Fig.~\ref{fig:loceff} shows the localization efficiency factor $H(T)$ as a function of temperature for different characteristic temperature scales $T_*=$ 200 MeV, 250 MeV, 300 MeV, 350 MeV, and 400 MeV for various exponent values $p=0.5, 1.0, 1.5$ and $2.0$\,. Observe that for a fixed value of $p$, the localization efficiency factor $H(T)$ decreases
as the temperature $T$ increases. For a fixed temperature $T$, localization efficiency is higher for larger values of $T_*$, since the term $\left(\dfrac{T}{T_*}\right)$ in the denominator of $H(T)$ is more suppressed. Furthermore the vanishing saturation value of $H(T)$ is reached earlier for larger values of the exponent $p$, as it is controlled by the factor $\dfrac{1}{T^p}$, that is for $T>>T_*$ 
\begin{align}
   H(T)= \dfrac{1}{1+\left(\dfrac{T}{T_*}\right)^p}
    \approx \left(\dfrac{T_*}{T}\right)^p\,. 
\end{align}
For $T<T_*$, the drop in $H(T)$ is steeper for smaller values of $p$, where stronger localization efficiency is favored by higher values of $p$. Once the temperature $T$ exceeds the characteristic temperature scale $T_*$, the drop is more pronounced for higher values of $p$, leading to more suppressed localization at higher values of $p$. Additionally, when  $T=T_*$, the value of the localization efficiency factor becomes $1/2$\,.

\begin{figure}[htbp]
    \centering
    \includegraphics[width=8.1cm]{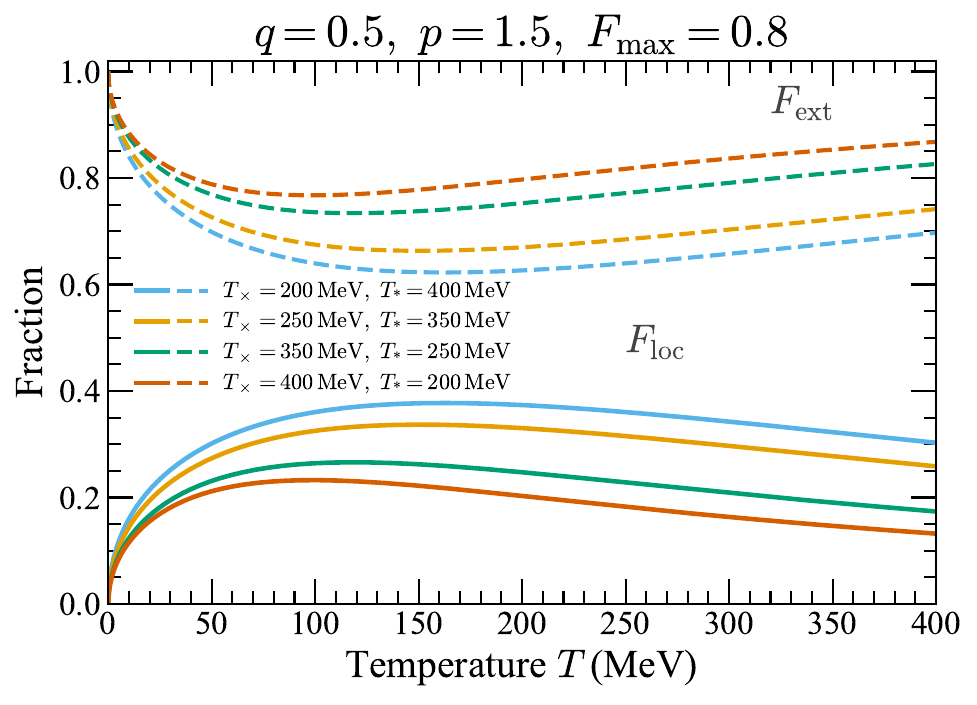}
    \includegraphics[width=8.1cm]{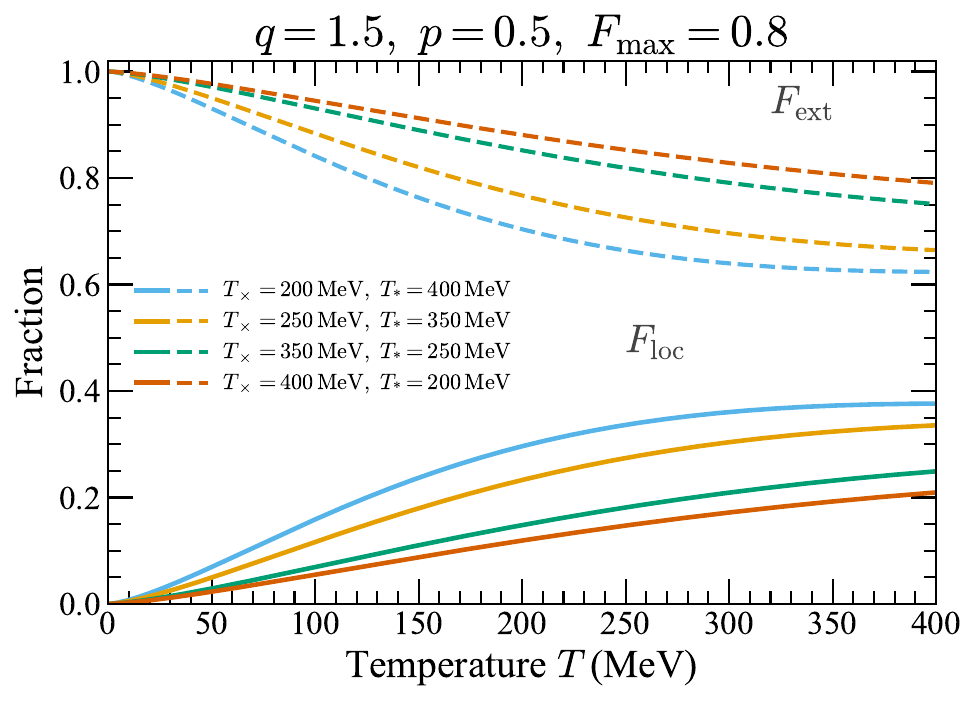} \\
     \includegraphics[width=8.1cm] {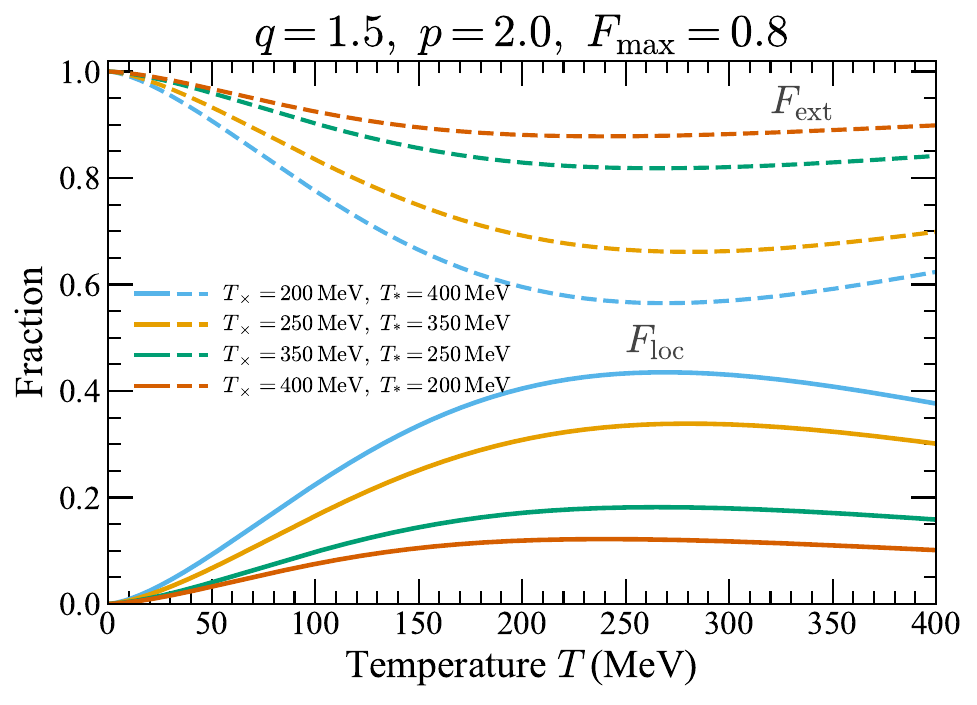}
    \includegraphics[width=8.1cm]{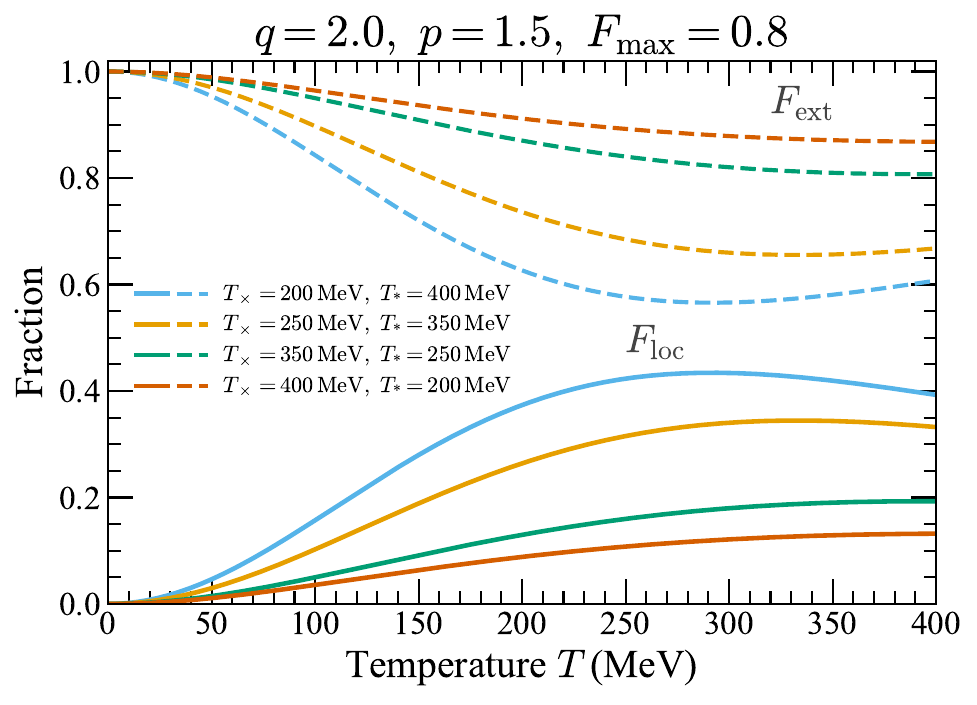}
    \caption{The plots of the localization and extended fractions $F_{loc}(T)$ and $F_{ext}(T)$ as functions of temperature for varying characteristic temperature scales $(T_\times,T_*)= (200\,\rm{MeV}, 400\,\rm{MeV}), (250\,\rm{MeV}, 350\,\rm{MeV}),(350\,\rm{MeV}, 250\,\rm{MeV}),\:\:\rm{and}\:\:(400\,\rm{MeV}, 200\,\rm{MeV}) $. The corresponding exponent pairs are $(q,p)=(1.5,2.0), (2.0, 1.5), (1.5, 0.5)$, and $(0.5, 1.5)$ with a cap constant $F_{max}=0.8$\,. }
    \label{fig:locextfraction} 
\end{figure}

The Fig.~\ref{fig:locextfraction} shows the localization and delocalization fractions $F_{loc}(T)$ and $F_{ext}(T)$ plotted for different pair of exponent values $(q,p)$ and characteristic temperatures $(T_\times, T_*)$ with a cap value of $F_{max}=0.8$. 
It is important to note that the sum of the dashed and solid lines corresponding to the same benchmark parameters satisfies the normalization condition $F_{loc}(T)+F_{ext}(T)=1$. Observe that the localization fraction $F_{loc}(T)$ gradually increases with temperature until it reaches its maximum value at a particular temperature $T_{(q,p)}^{max}$, where it begins to drop gradually. This temperature $T_{(q,p)}^{max}$ corresponds to the disorder that localizes the quarks most effectively, resulting in the maximum thermodynamic suppression. Since the extended fraction $F_{ext}(T)$ is complementary to the localization fraction $F_{loc}(T)$, the two fractions exhibit completely opposite behavior. For the pair of characteristic temperatures $(T_\times, T_*)=(200\,\rm{MeV}, 400\,\rm{MeV})$, the largest (smallest) localization (extended) fraction is obtained as seen in~Fig.~\ref{fig:locextfraction}\,. This occurs due to the condition $T_\times << T_*$ for maximal localization as reflected in Eq.~\eqref{eq:maximalcondition}\,. For lower temperatures $T<T_{(q,p)}^{max}$, a higher localization fraction is favored by a smaller characteristic temperature scale because 
\begin{align}
 G(T)H(T)\approx \left(\dfrac{T}{T_\times}\right)^q,    
\end{align}
as shown in Eq.~\eqref{eq:disorederactivationapprox} with $H(T)\approx 1$\,. Therefore, the localization fraction behaves as $F_{loc}(T)\propto {T}^q$, which becomes larger as temperature increases. In addition, the localization fraction becomes even bigger for higher values of $q$\,.
Now for $T>T_{(q,p)}^{max}$ with $T_\times < T$, the localization fraction $F_{loc}(T)$ decreases as temperature increases since the decay rate of the localization efficiency factor dominates in this regime. Here, the larger values of $T_*$ yield higher $F_{loc}(T)$ as seen in Eq.~\eqref{eq:localizationfactor}\,.

Moreover, by fixing the characteristic temperature scale $T_\times$, the localization fraction gradually increases with temperature for $T<<T_{(q,p)}^{max}$, and the localization fraction is larger for smaller values of $q$. Also, when $T=T_\times=T_*$ the value of $F_{loc}=(0.63)(0.5)=0.315$\,. The maximum (minimum) values of the localization (extended) fraction curves are listed in Table~\ref{tab:floc_fext_extrema_pair}\,.   

\begin{table}[h!]
\centering
\begin{tabular}{|c|c|c|c|}
\hline
\textbf{$(q,p)$} & {$(T_\times,\,T_*)$ [MeV]} & {$T_{(q,p)}^{\max}$ [MeV]} & {$(F_{\rm loc}^{\max},\,F_{\rm ext}^{\min})$} \\
\hline
(0.5,\,1.5) & (200,\,400) & 162 & (0.38, 0.62) \\
(0.5,\,1.5) & (250,\,350) & 151 & (0.34, 0.66) \\
(0.5,\,1.5) & (350,\,250) & 118 & (0.27, 0.73) \\
(0.5,\,1.5) & (400,\,200) &  99 & (0.23, 0.77) \\
\hline
(1.5,\,0.5) & (200,\,400) & 400 & (0.38, 0.62) \\
(1.5,\,0.5) & (250,\,350) & 400 & (0.34, 0.66) \\
(1.5,\,0.5) & (350,\,250) & 400 & (0.25, 0.75) \\
(1.5,\,0.5) & (400,\,200) & 400 & (0.21, 0.79) \\
\hline
(1.5,\,2.0) & (200,\,400) & 269 & (0.44, 0.56) \\
(1.5,\,2.0) & (250,\,350) & 280 & (0.34, 0.66) \\
(1.5,\,2.0) & (350,\,250) & 265 & (0.18, 0.82) \\
(1.5,\,2.0) & (400,\,200) & 240 & (0.12, 0.88) \\
\hline
(2.0,\,1.5) & (200,\,400) & 292 & (0.43, 0.57) \\
(2.0,\,1.5) & (250,\,350) & 333 & (0.34, 0.66) \\
(2.0,\,1.5) & (350,\,250) & 393 & (0.19, 0.81) \\
(2.0,\,1.5) & (400,\,200) & 400 & (0.13, 0.87) \\
\hline
\end{tabular}
\caption{Maximum localization and minimum extended fraction for different sets of parameter values shown in the plots of Fig.~\ref{fig:locextfraction} with upper bound temperature $T=400$\,\rm{MeV}.  }
\label{tab:floc_fext_extrema_pair}
\end{table}

\begin{figure}[htbp]
    \centering
    \includegraphics[width=8cm]{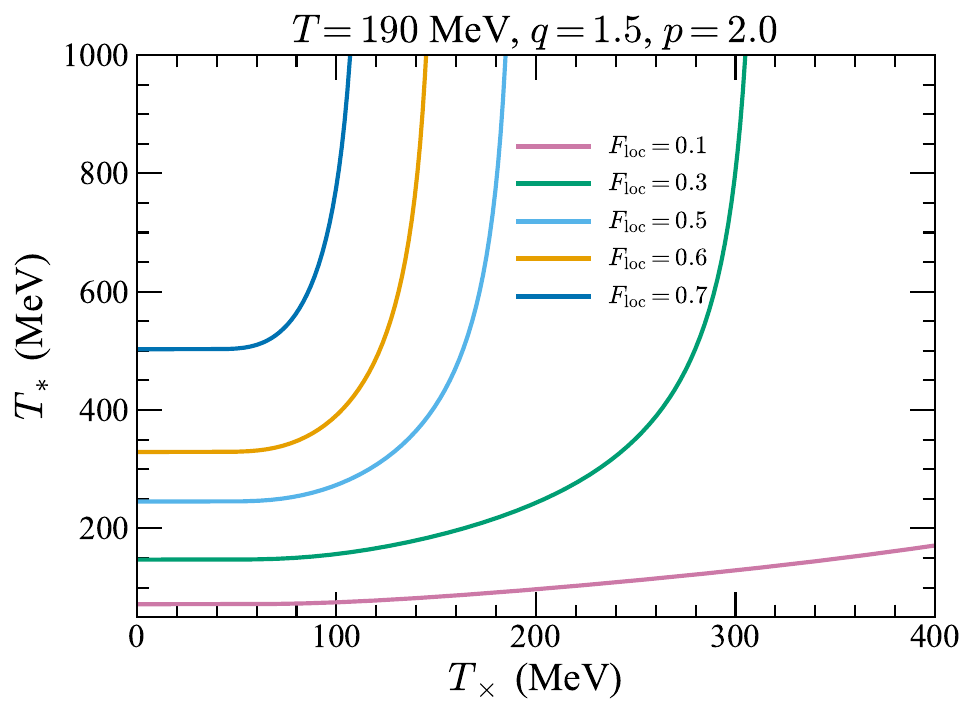}
    \includegraphics[width=8cm]{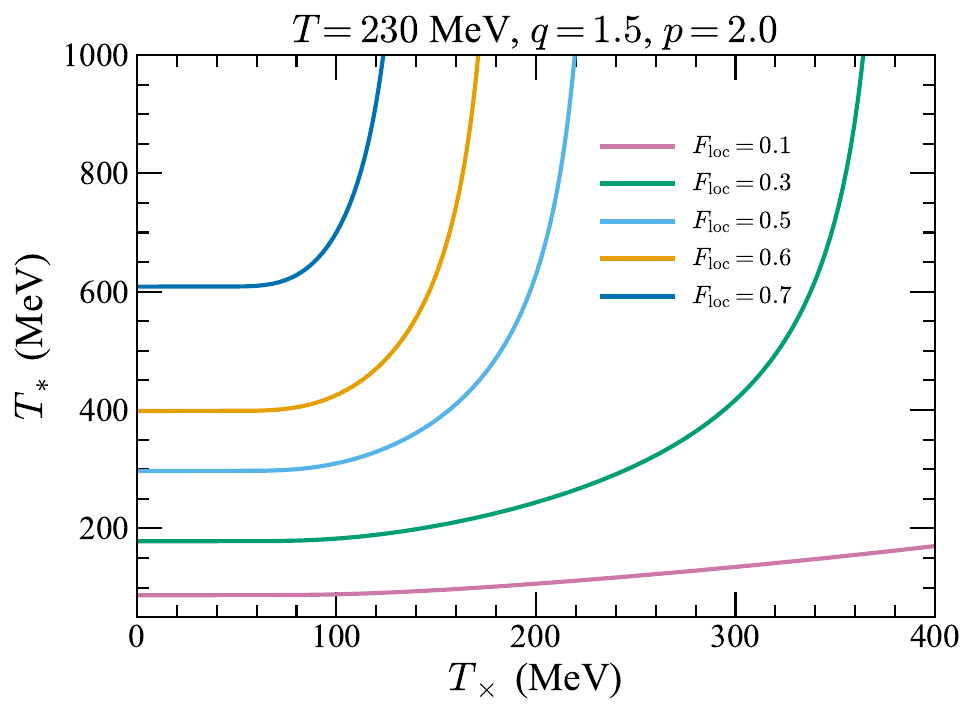} \\
     \includegraphics[width=8cm] {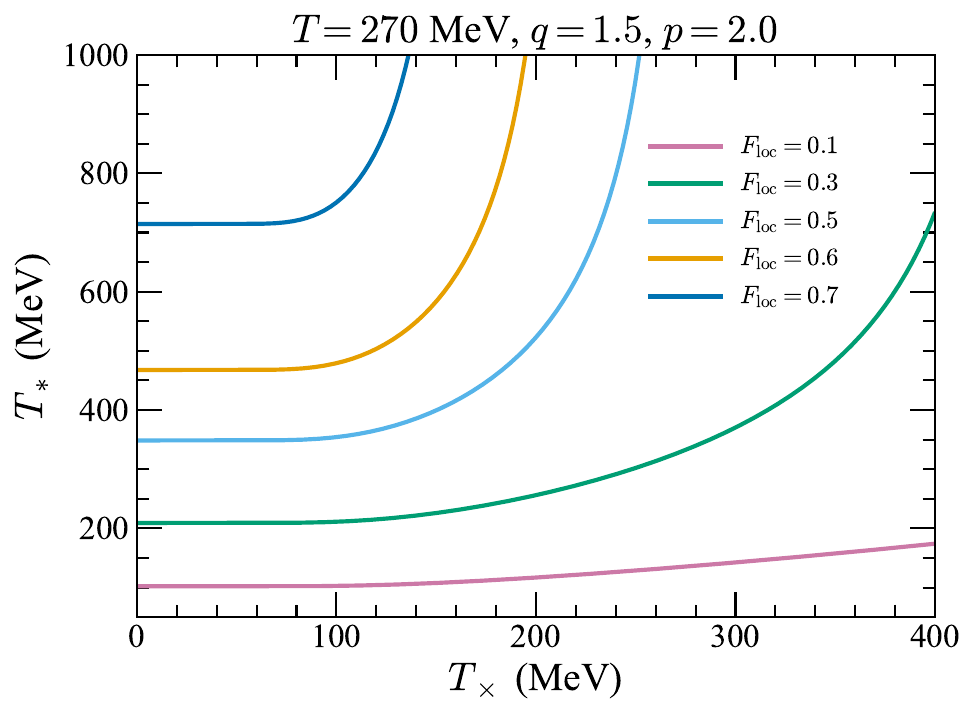}
    \includegraphics[width=8cm]{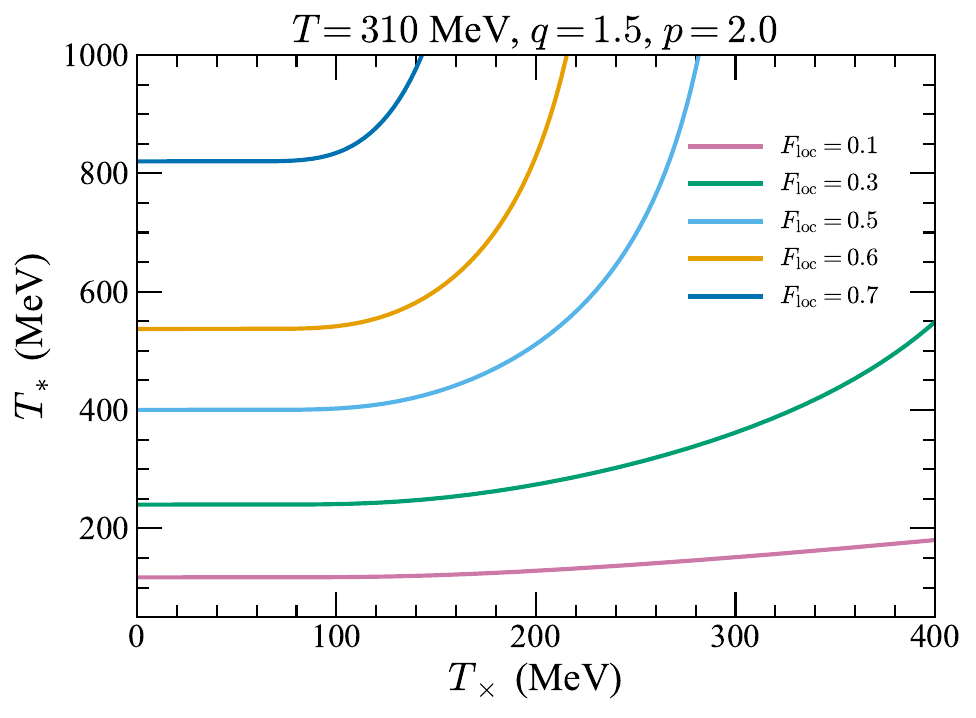}
    \caption{The figure shows the contour plots of the localization fraction $F_{loc}(T)=0.1,0.3, 0.5, 0.6$, and $0.7$ for different temperatures $T=190$ MeV, 230 MeV, 270 MeV, and 310 MeV with exponent values $q=1.5$ and $p=2.0$\,.}
    \label{fig:contourplotfloc} 
\end{figure}

The Fig.~\ref{fig:contourplotfloc} shows the contour plots of the localization fraction $F_{loc}(T)$ for the values $0.1, 0.3, 0.5, 0.6$, and $0.7$ with a pair of exponents $(q,p)=(1.5,2.0)$\,.
It can be observed that for a fixed temperature $T$, the localization fraction $F_{loc}(T)$ value increases as the contour curve shifts to the left. This behavior arises since 
the disorder activation function $G(T)$ is larger for smaller values of $T_\times$, while the localization efficiency factor $H(T)$ increases for larger values of $T_*$\,.  That is, the localization is favored by smaller $T_\times$ and larger $T_*$\,.

\subsection{Time evolution of EOS with Localization}
\begin{figure}[htbp]
    \centering
    \includegraphics[width=15cm]{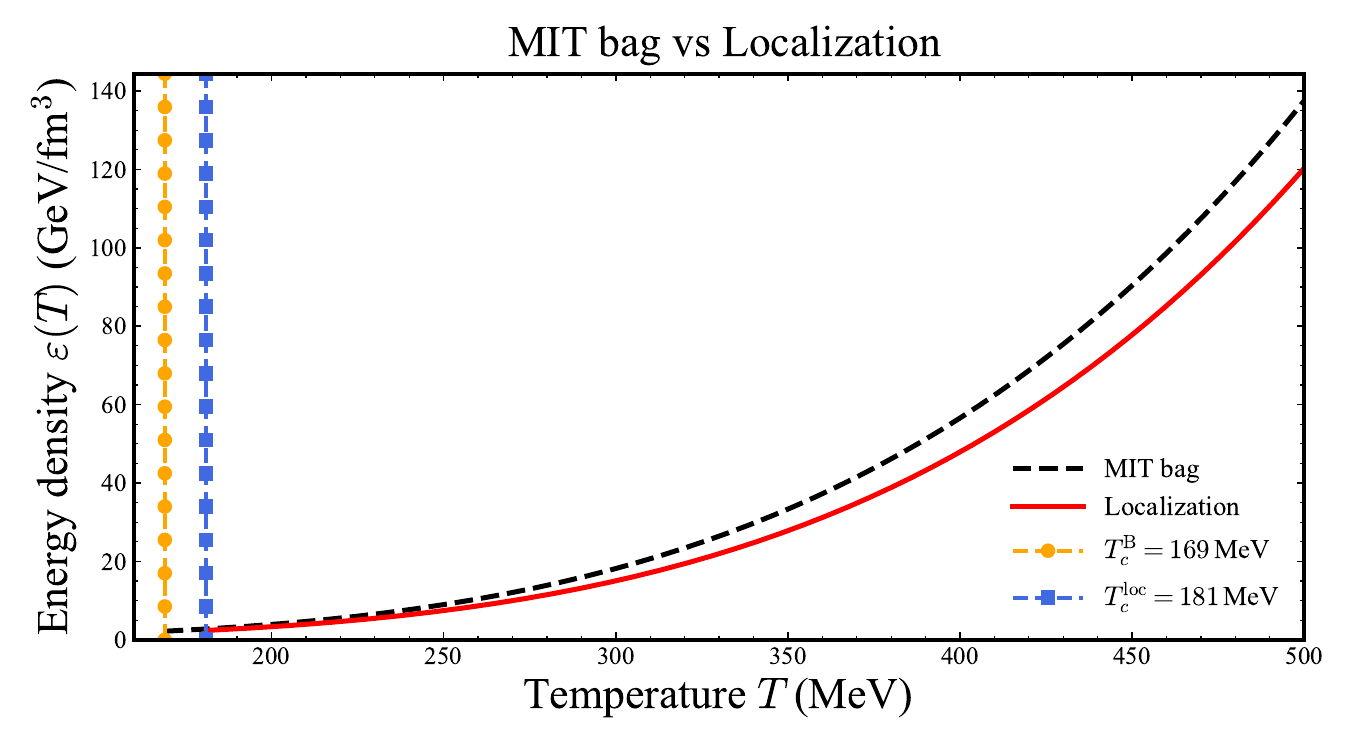}
     \caption{The temperature dependence of the MIT bag and localization model for two quark flavors. The red solid (black dashed) line shows the energy density of the localization (MIT bag) model.     
     The benchmark parameters used are  $B=235\,(\textrm{MeV})^4$, $T_\times=200\,\textrm{MeV}$, $T_*=400\,\textrm{MeV}$, $p=2.0$, $q=1.5$, and $F_{max}=0.8$\,.  }
    \label{fig:energydensityvstemperature} 
\end{figure}

Let us now analyze the time evolution of the modified bag model EOS with localization. The time evolution for the energy density and pressure can be determined from Eq.~\eqref{eq:Friedmanneq}\,. In our numerical calculations, we utilize the initial condition
for the temperature as $T_{loc}(t_0)=500\,\textrm{MeV}$ with $t_0=1.35\, \mu s$, which comes from the realistic QCD EOS applied to relativistic heavy-ion collisions at RHIC~\cite{Florkowski:2010mc}\,. Particularly, we apply it to the time evolution of the energy density and pressure for both the bag model and its phenomenological modification that includes quark localization. 
As shown in Fig.~\ref{fig:energydensityvstemperature}, the energy density is plotted as a function of temperature. The black dashed line denotes the bag model, while the red solid line represents the localization model. Also, the curves describe the scenario prior to the onset of phase transition, that is, when the temperature $T$ is above the critical temperature for each model. The benchmarks $T_\times=200$ and $T_*=400$ are used in the localization model.
As observed in Fig.~\ref{fig:energydensityvstemperature}, the energy density obtained from the localization model is lower than that of the bag model. The plots also indicate the critical temperature $T_{c}^{loc}= 181\, \textrm{MeV}$ for the localization model computed using Eq.~\eqref{eq:criticaltemploc}\,. This value is higher than the critical temperature of the bag model $T_c^{B}=169\,\textrm{MeV}$, corresponding to the bag constant $B=(235\,\textrm{MeV})^4$\,. In this case, the localization model raises the critical temperature $T_c^{B}$ by 7\,\%\,. 

\begin{figure}[htbp]
    \centering
    \includegraphics[width=9.5cm]{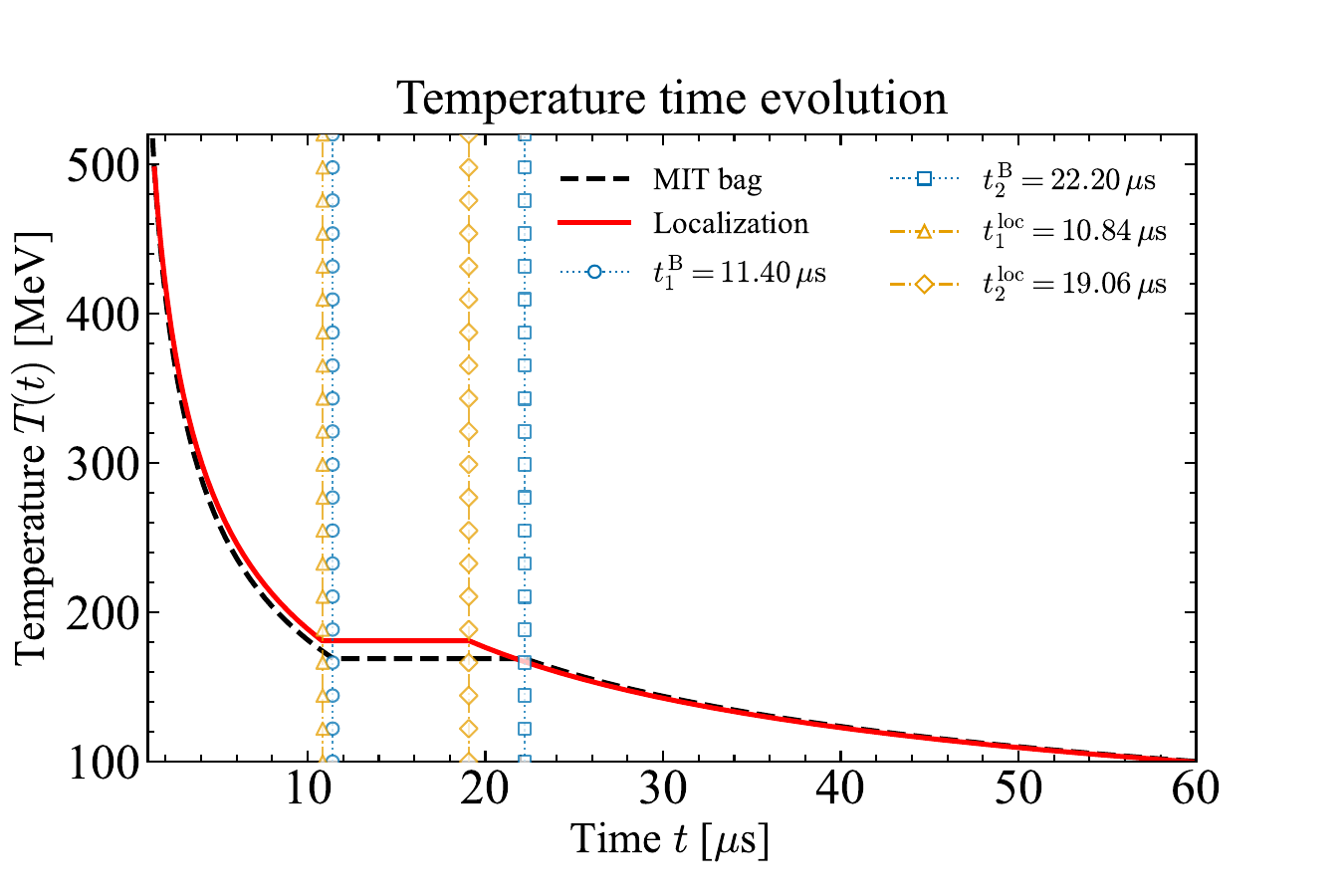}\\
    \includegraphics[width=10.5cm]
{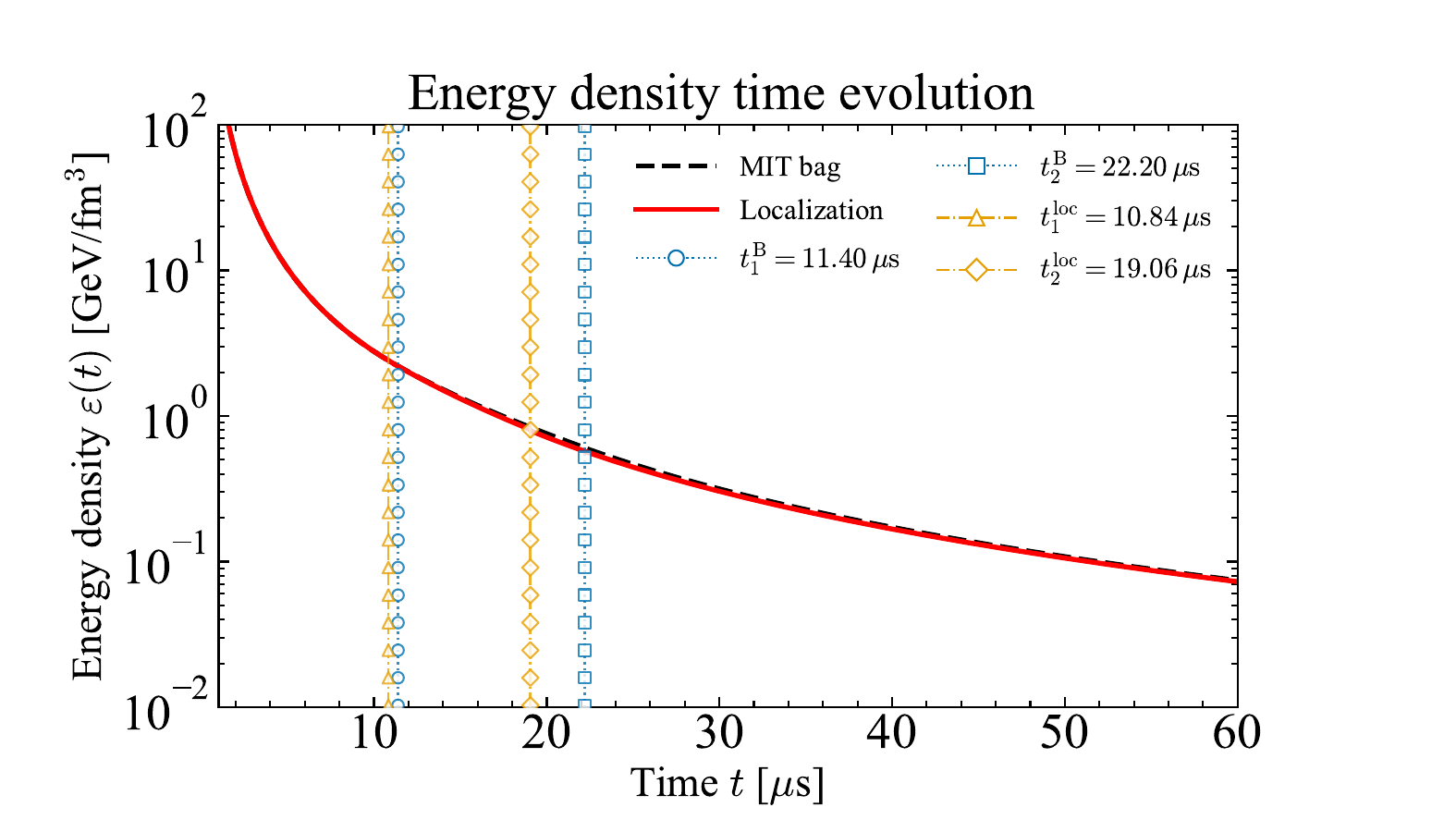}\\
    \includegraphics[width=10.5cm]{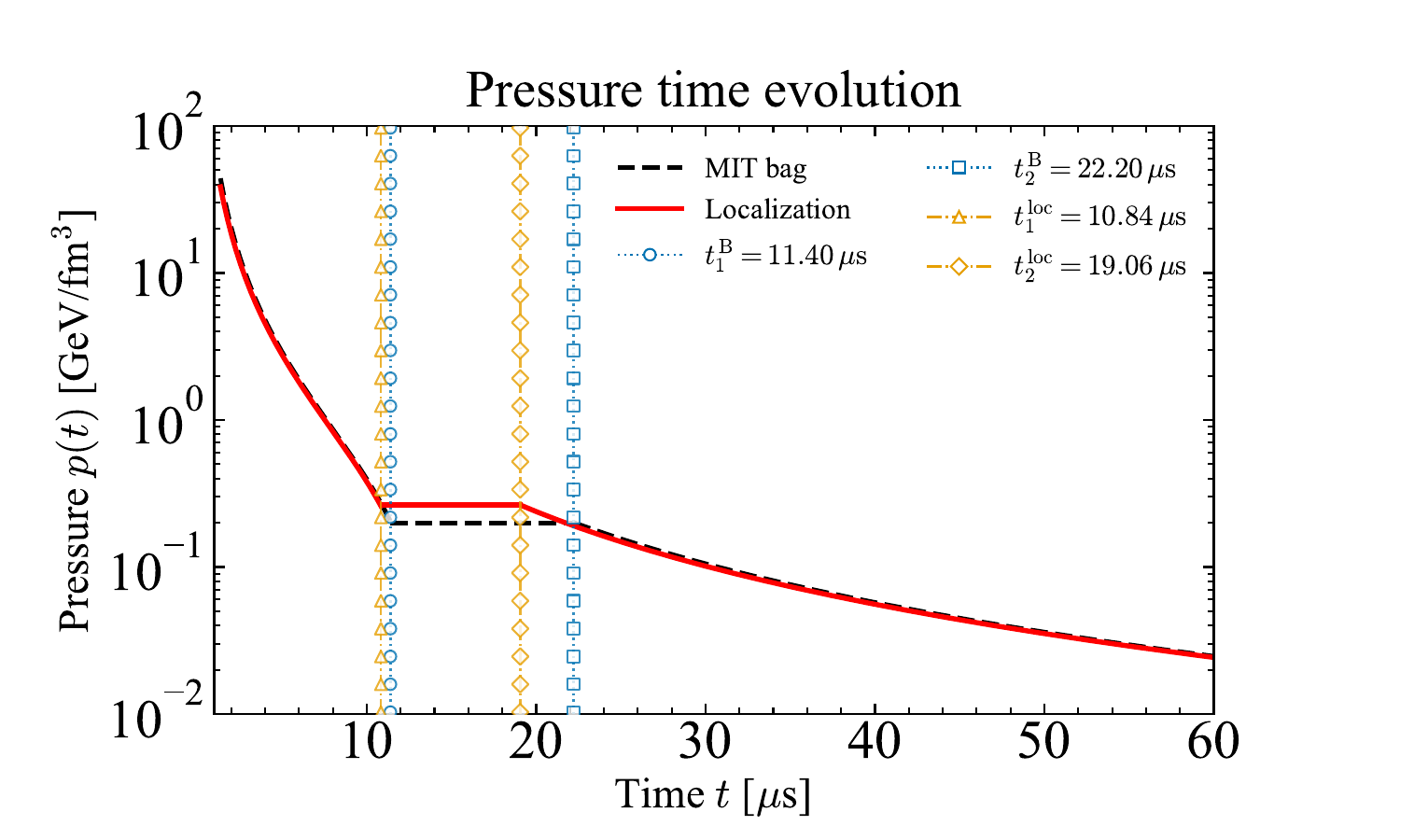}
     \caption{The time evolution of the temperature, energy density, and pressure of the MIT bag and localization model for two quark flavors. The red solid (black dashed) line represents the time evolution curves of the localization (MIT bag)  model. The benchmark parameters used are  $B=235\,(\textrm{MeV})^4$, $T_\times=200\,\textrm{MeV}$, $T_*=400\,\textrm{MeV}$, $p=2.0$, $q=1.5$, and $F_{max}=0.8$ with initial condition $T_{loc}(t_0)=T_B(t_0)=500\,\textrm{MeV}$ such that $t_0=1.35\,\mu\textrm{s}$\,. }
    \label{fig:TempEDandPandenergydensityvsTime} 
\end{figure}

In the upper panel of Fig.~\ref{fig:TempEDandPandenergydensityvsTime}, we can see the comparison of the time evolution of temperature of the bag model (black dashed line) and localization model (red solid line). As shown in this figure, the localization model shifts the bag model's temperature curve upward in the QGP phase. The onset of the phase transition now occurs earlier at time $t_1^{loc}=10.84\,\mu s$ and ends at $t_2^{loc}=19.06\, \mu s$, as compared to the bag model which transition from $t_{1}^B=11.40\,\mu s$ to $t_2^{B}=22.2\,\mu s$, as determined from  Eq.~\eqref{eq:t1localization}\,. This is a consequence of the reduced d.o.f. in the localization model, which causes hadronization to begin at higher temperature.
In addition, the time interval from the onset of the phase transition to its completion becomes shorter, given by $\Delta t^{loc}=t_{2}^{loc}-t_{1}^{loc}=8.22\,\mu s$. However, for the bag model this corresponds to $\Delta t^B=t_{2}^B-t_{1}^B=10.80\, \mu s$ as calculated using Eq.~\eqref{eq:timeintervalphasetransition}. In this case, the localization shortens the duration of the mixed phase by roughly 24\,\%. This reduction is reflected in the smaller value of $\sqrt{s-1}$ in Eq.~\eqref{eq:timeintervalphasetransition} with $s=\dfrac{g^{eff}}{g_2}$\,. Remarkably in the hadronic phase, after time $t_{2}^B=22.20 \,\mu s$, the temperature in the localization model becomes lower than that of the bag model. This means that localization of quarks in the QGP phase leads to a faster cooling of the hadronic phase for time $t> t_{2}^B$\,. 

The time evolution of the energy density in the localization model reflected in the uppermost panel of Fig.~\ref{fig:TempEDandPandenergydensityvsTime} is slightly lower than that of the bag model. Furthermore, the pressure in the lowermost panel of Fig.~\ref{fig:TempEDandPandenergydensityvsTime} is slightly lower than the bag model for time $t< t_{1}^{loc}$ and $t> t_{2}^{B}$, and it is larger during the mixed phase transition. For the localization model, at the onset of the phase transition at time ($t=t_{1}^{loc}=10.84 \,\mu s <t_{1}^B=11.40\, \mu s$), the critical pressure value is $p_{c}^{loc}=0.265\,GeV/fm^3$. It gradually decreases until time $t_{2}^B=22.20 \,\mu s$, where it becomes lower than the critical pressure of the bag model with $p_{c}^B=0.200 \,GeV/fm^3$\,. This means that the critical pressure $p_{c}^{loc}$ in the localization model is reached earlier as the system evolves from QGP gas phase. That is, the equality of pressure in the plasma and hadronic phase occurs sooner, compensating for the reduced d.o.f. in QGP as evident by the critical temperature in Eq.~\eqref{eq:localozationtemperature} where $T_{c}^{loc}>T_c^B$ due to the suppressed value of $g_{q}^{eff}$
\,. 

\begin{figure}[htbp]
    \centering
    \includegraphics[width=11cm]{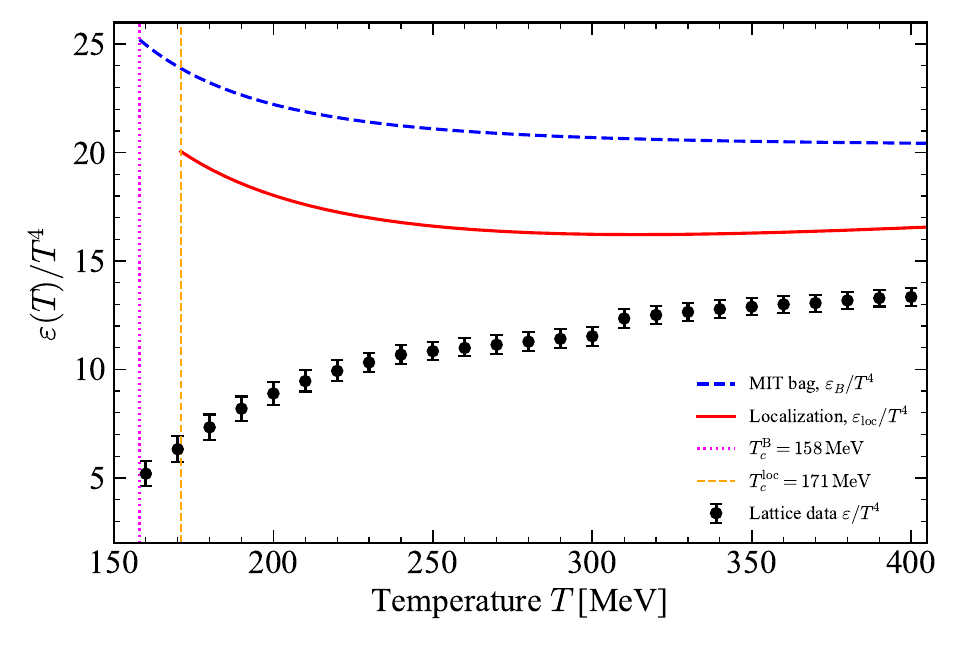}
    \includegraphics[width=11cm]
    {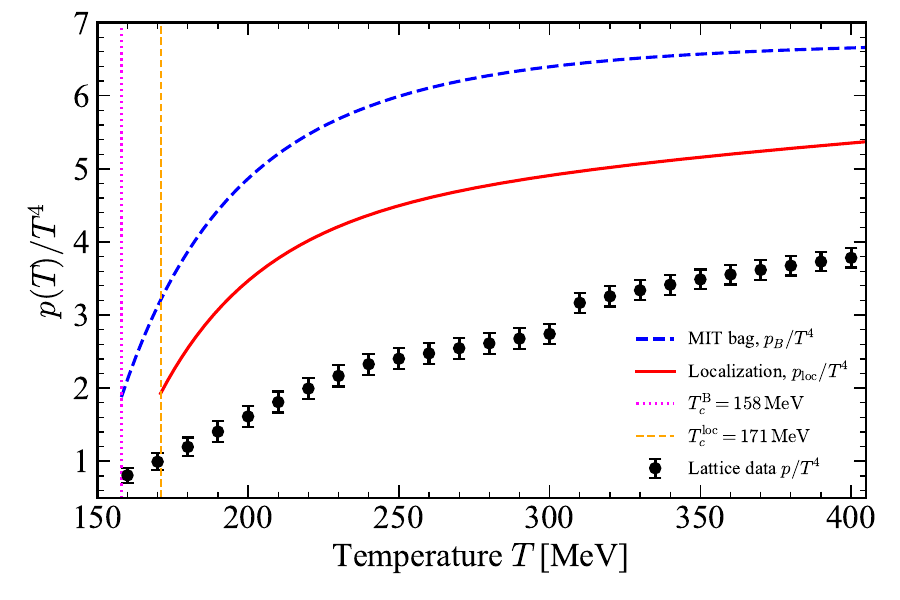}
     \caption{The temperature dependence of the MIT bag and localization model above their critical temperature for $(2+1)$ quark flavors. The red solid (blue dashed) line shows the energy density and pressure of the localization (bag) model, respectively. The lattice data indicated by black points with error bars are taken from the HotQCD collaboration~\cite{HotQCD:2014kol}. The benchmark parameters used are  $B=235\,(\textrm{MeV})^4$, $T_\times=200\,\textrm{MeV}$, $T_*=400\,\textrm{MeV}$, $p=2.0$, $q=1.5$, and $F_{max}=0.8$ with initial condition $T_{loc}(t_0)=T_B(t_0)=500\,\textrm{MeV}$ such that $t_0=1.35\,\mu\textrm{s}$\,. }
    \label{fig:energydensityandpressuret4vsTemperature} 
\end{figure}

The plots in Fig.~\ref{fig:energydensityandpressuret4vsTemperature} show the temperature dependence of the energy density and pressure for the bag model (blue dashed line) and the localization model (red solid line), respectively. The points with the error bars represent the lattice QCD data from the  HotQCD collaboration~\cite{HotQCD:2014kol}\,. In both panels in Fig.~\ref{fig:energydensityandpressuret4vsTemperature}, we see that the trend of the localization model is
closer to the lattice data than the MIT bag model. As shown in the upper panel of Fig.~\ref{fig:energydensityandpressuret4vsTemperature}, in the the low temperature regime $T>T_c^{loc}$, the ratio $\dfrac{\varepsilon_{loc}({T})}{T^4}$ decreases gradually with increasing temperature. This behavior arises from the diminishing contribution of the non-negligible term $\dfrac{B}{T^4}$, while the values get closer to the lattice data. It also approaches the reduced Stefan-Boltzmann limit $g^{eff}\dfrac{\pi^2}{30}$ for very high temperature $T$ where it best describes the lattice data, as obtained from Eq.~\eqref{eq:localizationenergydensityandpressure} since $\dfrac{B}{T^4}<<1$\,. The Stefan-Boltzmann limit discrepancy between the bag model and localization model is given by $\Delta g\dfrac{\pi^2}{30}$ where $\Delta g=(g_{qgp}+g_{ew}-g^{eff})$\,. Essentially, the difference $\Delta g$ in the internal d.o.f. of the QGP reflects the effects of quark localization, which decreases their thermodynamic participation resulting to lower energy density. The lower panel of Fig.~\ref{fig:energydensityandpressuret4vsTemperature} shows the relationship between the ratio $p(T)/T^4$ and temperature $T$\,. The localization model ratio $\dfrac{p_{loc}(T)}{T^4}$ moves to higher values as the temperature $T$ increases since the term $-\dfrac{B}{T^4}$ is becoming less negative. The localization model exhibits the same tendency as the lattice data when the temperature increases. In addition for very high temperature $T$, the term $-\dfrac{B}{T^4}$ becomes negligible causing the ratio $\dfrac{p_{loc}(T)}{T^4}$ to approach the value $g^{eff}\dfrac{\pi^2}{90}$ with pressure discrepancy from the bag model given by $\Delta g \dfrac{\pi^2}{90}$\,. Consequently, the reduction in pressure value $p_{loc}(T)$ in the localization model makes the onset of mixed phase duration to occur at higher temperature and earlier times in the early-universe scenario\,. Overall, we can conclude that the localization model provides a significantly better description of the lattice data than the bag model. 

\section{Conclusion}
\label{section:conclusion}
The MIT bag model provides a classic description of the EOS that portrays the thermodynamic behavior of the quark-gluon plasma and hadronic phase. We extended this model by including the electroweak sector in the total energy density  and pressure, and introducing quark localization effects in the QGP phase.  Notably, the inclusion of the former that contributes additional d.o.f. does not alter the critical temperature $T_{c}^B$ of the phase transition in the bag model. Also, the choice of the bag constant $B=(235\,\textrm{MeV})^4$ gives the corresponding critical temperature $T_c^B=169\,\textrm{MeV}$ for the bag model with two quark flavors. The latter, on the other hand, results in a temperature-dependent effective number of quarks d.o.f. $g_{q}^{eff}=F_{ext}(T)g_q$\,. The suppression factor $F_{ext}(T)$ arises from the localization efficiency induced by the disorder from the gluon fields. This yields an increase of the critical temperature  roughly by 7\%\,.

Utilizing the Friedmann equation for the localization model, the onset of the phase transition occurs earlier at $t_1^{loc}=10.84\,\mu s$ at critical temperature $T_c^{loc}=181\,\textrm{MeV}$. The time interval for the mixed phase duration $\Delta t_{loc}=8.22\, \mu s$ is shortened by about 24\% in comparison with the bag model, which is $\Delta t_{B}=10.80\, \mu s$\,. Consequently, the hadronic phase cools faster in this case. The equality of pressure between QGP and hadronic phases occurs earlier, compensating for the reduced quarks' d.o.f.\,. The pressure during the mixed phase duration is also higher than the bag model due to the increased critical temperature $T_{c}^{loc}$\,.  We have shown that the localization model is in better agreement using the lattice data from the HotQCD collaboration. Moreover, the effective Stefan-Boltzmann limit is reached at much higher temperature, significantly shifting the bag model curves closer to the lattice data in the high temperature regime.

\section*{Acknowledgments}
We gratefully acknowledge the support of the Institute for Basic Science – Center for Theoretical Physics of the Universe (IBS-CTPU) in Daejeon, South Korea. In particular, we extend our sincere thanks to Prof. Kiwoon Choi, who provided the opportunity for a research visit at IBS-CTPU.

\end{document}